\documentclass[showpacs,preprintnumbers,prd,nofootinbib,floats,amssymb,floatfix]{revtex4}
\usepackage{amsmath}
\usepackage{amsfonts}
\usepackage{graphicx}
\usepackage{amsxtra}
\usepackage{eufrak}
\usepackage{hyperref}
\usepackage{amsmath}
\usepackage{amstext}
 
\begin{document}
\title{ Canonical and   symplectic analysis   for three dimensional gravity without dynamics}
\author{Alberto Escalante}  \email{aescalan@ifuap.buap.mx}
 \affiliation{  Instituto de F{\'i}sica, Benem\'erita Universidad Aut\'onoma de Puebla, \\
 Apartado Postal J-48 72570, Puebla Pue., M\'exico, }
\author{ H. Osmart Ochoa-Guti\'errez }  
 \affiliation{ Facultad de Ciencias F\'{\i}sico Matem\'{a}ticas, Benem\'erita Universidad Au\-t\'o\-no\-ma de Puebla,
 Apartado postal 1152, 72001 Puebla, Pue., M\'exico.}
\begin{abstract}
In this paper a detailed Hamiltonian analysis of three-dimensional gravity without dynamics proposed  by V. Hussain  is performed. We report  the complete structure of the constraints and the Dirac brackets are explicitly computed. In addition,  the  Faddeev-Jackiw symplectic approach is developed;  we report  the complete set of Faddeev-Jackiw constraints and  the generalized   brackets, then we show  that the Dirac and the generalized Faddeev-Jackiw brackets coincide to each other.   Finally, the similarities and advantages between  Faddeev-Jackiw  and   Dirac's formalism   are briefly discussed.
\end{abstract}
 \date{\today}
\pacs{98.80.-k,98.80.Cq}
\preprint{}
\maketitle
\section{Introduction}
It is well-known that the  quantum study of gravity developed in Loop Quantum Gravity [LQG]  is based on a background independent   and non-perturbative  canonical formulation \cite{1, 2, 3, 3aa, 3a4, 3a4a}. In fact, the support of LQG is the canonical quantization scheme  developed by  Dirac-Bergman \cite{4a}.  Dirac's canonical formalism is a powerful  approach, it allows us identify the physical degrees of freedom, the gauge transformations, the
complete structure of the constraints and the obtention of the extended action,  this information
is useful because a strict study of the symmetries will allow us to have  the best  guideline to make   the quantization. Nonetheless, if a  complete Dirac's canonical analysis is performed, in general it is
complicated to classify  the constraints in first and second class; the classification
of the constraints is an important step to perform because first class constraints are generators
of gauge transformations and allow us identify observables and second class constraints allow
us to construct the Dirac brackets. However,  in spite of  Dirac's framework is a powerful tool for analyzing constrained systems, the quantum  canonical formulation of gravity has been  a difficult task to perform. For that reason,  the study of  toy models  with a similar canonical structure  to  that  present in  gravity  becomes to be an  interesting topic. In this respect, there are several examples of toy  models  with a close relation with  gravity just as topological theories \cite{4, 5, 6, 7, 8, 9, 10, 11, 12}  and models as those   reported by  V. Hussain where there is not dynamics \cite{13, 14}, this means, in the canonical formulation of those  models there is  not   an analog to the Hamiltonian constraint  that is present  in  real gravity theory \cite{15}. In this respect, the study of toy models is a interesting  subject  because those are good laboratories for testing classical and quantum ideas  that could be applied to  General Relativity [GR]. \\
Because of the explained  above, in this paper the Faddeev-Jackiw   analysis  for  three dimensional gravity without  dynamics   reported in \cite{13} is performed. In fact, the FJ framework is an alternative approach for studying constrained systems \cite{15, 16, 17},  the degrees of freedom are identified by means the so-called symplectic variables and these variables allow us to construct a symplectic matrix. In this manner,  in FJ scheme all relevant information is contained in the symplectic matrix.  Furthermore, since the system under study is singular there will be constraints and the FJ approach has the advantage that all the constraints of the theory are at the same footing, namely, it is not necessary perform the classification of the constraints in primary, secondary, first class or second class such as in Dirac's method is done. Moreover, in  FJ approach also it is possible to obtain the gauge transformations of the theory, and by fixing the gauge a symplectic tensor is constructed, then from that symplectic tensor it is possible to identify   the generalized FJ brackets;  generalized FJ and Dirac's brackets  coincide  to each other. However,  just as it  has been commented in  \cite{18}  in order to compare the   results of   Dirac's approach with the  FJ  ones, it is necessary develop  a complete Dirac's analysis, this is, it is mandatory to follow all Dirac's steps. In this respect, in this paper  we  develop  a complete Dirac's analysis of the theory under study, then we  compare the Dirac  results with the FJ ones; we will conclude that the FJ framework  is more economical  than Dirac's  scheme.\\
The paper is organized as follows, in Section II a pure Dirac's analysis of three dimensional gravity without dynamics is performed. We identify the complete structure of the constraints and the second class constraints are eliminated by introducing the Dirac brackets.  In Section III we develop a complete FJ symplectic analysis of the theory under study. The FJ constraints and the symmetries of the theory jus as the gauge transformations are identified, then by fixing the gauge a symplectic tensor tensor  is constructed. From the symplectic tensor  the generalized FJ brackets are identified and we compare these brackets with the Dirac ones; we will show that the FJ and Dirac's brackets coincide to each other. Finally we present the conclusions and prospects.  

%%%%%%%%%%%%%%%%%%%%%%%%%%%%%%%%%%
\section{Hamiltonian analysis }
In this section a detailed canonical analysis for Husain's model is performed. As we have commented previously if we wish  compare the  canonical approach with the symplectic method, then it is necessary to perform the Hamiltonian analysis by following all Dirac's steps. In this manner, the action of interest  is given by \cite{13}
\begin{eqnarray}
S[e, \omega]=\int_{\mathcal{M}} \epsilon^{\alpha \beta \gamma} \left[   \omega_\alpha \partial_\beta \omega_\gamma  + \lambda e^{i}_\alpha \partial_\beta e^{i}_\gamma  + \lambda \epsilon_{ij} e^{i}_\alpha  e^{j}_ \beta \omega_\gamma  \right].
\label{1}
\end{eqnarray}
where $e^{i}_\alpha$ is the zweibein representing the gravitational field and $\omega_\alpha$ is a gauge field, $\lambda $ is the cosmological constant that can be positive or  negative (in this work we shall assume that $\lambda$ is positive),  $x^{\mu}$ are the coordinates that label the points of the 3-dimensional manifold $\mathcal{M}$. In our notation, Greek letters are indices for the spacetime and run from 0 to 2, while the middle alphabet letters  $(i,j,k=1,2)$   are associated with  the internal group and it can be raised and lowered with  the metric $\eta^{ij}$ with signature $(+, +)$. 
Thus, assuming that the manifold  $\mathcal{M}$ is topologically $\Sigma\times R$,  the 2+1 decomposition allow us identify the following  Lagrangian density 
\begin{eqnarray}
{\mathcal{L}}&=&\epsilon^{0ab}\omega_{b}\dot{\omega}_{a} + \lambda\epsilon^{0ab}e^{i}_{b}\dot{e}_{ai} + \omega_{0}\{2\epsilon^{0ab}\partial_{a}\omega_{b} + \lambda\epsilon^{0ab}\epsilon_{ij}e^{i}_{a}e^{j}_{b}\} + e^{i}_{0}\{2\lambda\epsilon^{0ab}\partial_{a}e_{bi} + 2\lambda\epsilon^{0ab}\epsilon_{ij}e^{j}_{a}\omega_{b}\}, \nonumber \\
\label{lag}
\end{eqnarray}
where $a,b,e = 1,2$. The matrix elements of the Hessian given by 
\begin{align}  \frac{\partial^{2}{\mathcal{L}}}{\partial(\partial_{\mu}e^{i}_{\alpha})\partial(\partial_{\mu}e^{i}_{\beta})},& &
\frac{\partial^{2}{\mathcal{L}}}{\partial(\partial_{\mu}e^{i}_{\alpha})\partial(\partial_{\mu}\omega_{\beta})},& &
\frac{\partial^{2}{\mathcal{L}}}{\partial(\partial_{\mu}\omega_{\alpha})\partial(\partial_{\mu }\omega_{\beta})},
\end{align}
are identically zero, thus, we expect 9 primary constraints. In order to identify the constraints,  the canonical formalism calls  for  the definition of the momenta $\left(\Pi^{a}_{i}, \Pi^{0}_{i},P^{0},P^{a}\right)$ canonically conjugate to $\left(e^{i}_{a},e^{i}_{0},\omega_{0},\omega_{a}\right)$ are given by
\begin{align}
 \Pi^{\alpha}_{i} &=\frac{\delta{\mathcal{L}}}{\delta \dot{e}^{i}_{\alpha}}, &
P^{\alpha} = \frac{\delta{\mathcal{L}}}{\delta \dot{\omega}_{\alpha}}.
\end{align}
From the null vectors of the Hessian and the  definition of the momenta, we identify the following 9 primary constraints
\begin{eqnarray}
\phi^{a}_{i} &:& \Pi^{a}_{i} - \lambda\epsilon^{0ab}e_{bi}\approx 0, \nonumber \\
\phi^{a} &:& P^{a} - \epsilon^{0ab}\omega_{b}\approx 0, \nonumber \\
\phi^{0}_{i} &:& \Pi^{0}_{i} \approx 0, \nonumber \\
\phi^{0} &:& P^{0}\approx 0.
\end{eqnarray}
In this manner, the canonical Hamiltonian takes the form
\begin{eqnarray}
{\mathcal{H}}_{c}&=&\int \{e^{i}_{0}\left(2\partial_{a}\Pi^{a}_{i} + 2\lambda\epsilon_{ij}e^{j}_{a}P^{a}\right) + \omega_{0}\left(2\partial_{a}P^{a} + \epsilon_{ij}e^{i}_{a}\Pi^{aj}\right)\}dx^{2},
\end{eqnarray}
and the primary Hamiltonian is defined as 
\begin{eqnarray}
{\mathcal{H}}_{p}&=& {\mathcal{H}}_{c} + \int \left[\lambda^{i}_{a}\phi^{a}_{i} + \lambda_{a}\phi^{a} + u^{i}_{0}\phi^{0}_{i} + u_{0}\phi^{0}\right]dx^{2}, 
\end{eqnarray}
where $\lambda^{i}_{a},\lambda_{a},u^{i}_{0},u_{0}$ are Lagrange multipliers enforcing the primary constraints. For this theory, the fundamental Poisson brackets of the theory are given by
\begin{eqnarray}
\{e^{i}_{\alpha}(x), \Pi^{\beta}_{j}(y)\} &=& \delta^{i}_{j}\delta^{\beta}_{\alpha}\delta^{2}(x-y),\nonumber  \\
\{w_{\alpha}(x), P^{\alpha}(y)\} &=& \delta^{\alpha}_{\beta}\delta^{2}(x-y).
\end{eqnarray}
In order to observe if there are more constraints, we calculate consistency of the constraints and we obtain the following 3 secondary constraints
\begin{eqnarray}
\psi_{i}&:& \partial_{a}\Pi^{a}_{i} + \lambda\epsilon_{ij}e^{j}_{a}P^{a} \approx 0,\nonumber  \\
\psi &:& \partial_{a}P^{a} + \frac{1}{2}\epsilon_{ij}e^{i}_{a}\Pi^{aj} \approx 0.
\end{eqnarray}
Consistency requires conservation in time of the secundary constraints, however, for this theory there are not third constraints. Now we need to classify  all the constraints in first class and second class. For this aim, we calculate the following 12 x 12
matrix whose entries are the Poisson brackets between the primary and secondary constraints, the non-zero brackets are given by 
\begin{eqnarray}
\{\phi^{a}_{i},\phi^{g}_{i}\} &=& -2\epsilon^{ag}\eta_{ij}\delta^{2}(x-y), \nonumber \\
\{\phi^{a}_{i}(x),\psi_{j}\} &=& \lambda\epsilon_{ij}P^{a} + \lambda\epsilon^{ag}\eta_{ij} \partial_{g}\delta^{2}(x-y),\nonumber  \\
\{\phi^{a}(x),\phi^{b}(y)\} &=& -2\epsilon^{ab}\delta^{2}(x-y), \nonumber \\
\{\phi^{a}(x),\psi_{j}\} &=& -\lambda\epsilon^{ag}\epsilon_{jk}e^{k}_{g}\delta^{2}(x-y), \nonumber \\
\{\phi^{a},\psi(y)\} &=& \epsilon^{ag}\partial_{g}\delta^{2}(x-y),
\end{eqnarray}
this matrix has $rank=6$ and 6 null vectors. Thus we expect 6 first class constraints and 6 second class constraints. Under a complicated work, we find the complete structure of the first class constraints  given by
\begin{eqnarray}
\phi^{0}_{i} &:&\Pi^{0}_{i} \approx 0, \nonumber \\
\phi^{0} &:& P^{0} \approx 0,\nonumber  \\
\gamma_{i} &=&  \partial_{a}\Pi^{a}_{i} + \lambda\epsilon_{ij}e^{j}_{a}P^{a} + \frac{1}{2}\epsilon_{ad}\epsilon^{l}_{i}P^{d}\phi^{a}_{l} - \frac{1}{2}\partial_{a}\phi^{a}_{i} - \frac{\lambda}{2}\epsilon_{ij}e^{j}_{b}\phi^{b} \approx 0, \nonumber  \\ 
\gamma &=& \partial_{a}P^{a} + \frac{1}{2}\epsilon_{ij}e^{i}_{a}\Pi^{aj} - \frac{1}{4}\epsilon_{ad}\epsilon^{li}\Pi^{d}_{l}\phi^{a}_{i} + \frac{\lambda}{4}\epsilon^{li}e_{la}\phi^{a}_{i} - \frac{1}{2}\partial_{b}\phi^{b} \approx 0,
\label{Fc}
\end{eqnarray}
where we can observe that $\gamma$ is the equivalent to the Gauss constraint and $\gamma_i$ is a  Gauss-like  constraint, there is not the analog to the Hamiltonian constraint that is  present in gravity, in this sense there is not dynamics. In fact,  $\gamma$ generates abelian transformations on the $\omega_a$ field and rotations on the  $e^i_\alpha$ field, we will observe this point  below. Furthermore, there are the following  second class constraints given by
\begin{eqnarray}
\chi^{a}_{i} &:& \Pi^{a}_{i} - \lambda\epsilon^{0ab}e_{bi} \approx 0, \nonumber \\
\chi^{a} &:& P^{a} - \epsilon^{0ab}\omega_{b} \approx 0 .
\end{eqnarray}
In this manner, the algebra between the constraints is given by 
\begin{eqnarray}
\{ \gamma_i(x), \gamma_j(y)\} &=& \frac{\lambda}{2} \epsilon_{ij} \gamma \delta^{2}(x-y),\nonumber  \\
\{ \gamma_i(x), \gamma (y)\}&=& - \frac{1}{2}\epsilon_{il} \gamma^l \delta^{2}(x-y),\nonumber  \\
\{ \gamma (x), \gamma(y)\} &=& 0,\nonumber  \\
\{ \gamma_i (x), \chi^{a} (y)\} &=& \frac{1}{2} \epsilon_{i}{^{l}} \chi^{a}_{l},\nonumber  \\
\{ \gamma_i (x), \chi^{a}_{j} (y)\} &=& \frac{\lambda}{2} \epsilon_{i j}\chi^{a},
\label{alg}
\end{eqnarray}
where we can see that the algebra between the constraints is closed as expected. It is important to comment that the identification of the   structure of the constraints (\ref{Fc}) is a difficult task to perform and it  has not been reported in the literature.  On the other hand,  with the information obtained until now, we can construct the Dirac brackets. For this aim we shall construct the matrix whose elements are only the Poisson brackets between the second class constraints, namely, $C_{\alpha\beta}(u,v) = \{\chi^{\alpha}(x), \chi^{\beta}(y)\}$ 

\begin{eqnarray}
\label{eq}
C_{\alpha \beta}(u,v)=
\left(
  \begin{array}{cccc}
   -2\lambda\epsilon^{ag}\eta_{il}  & \quad  0                                                                           \\
    0  &\quad    -2\epsilon^{ab}                                                                        \\
      \end{array}
\right)\delta^{2}(u-v),
\end{eqnarray}

and its inverse 

\begin{eqnarray}
\label{eq}
C^{-1}_{\alpha \beta}(u,v)=
\left(
  \begin{array}{cccc}
   \frac{1}{2\lambda}\epsilon_{ag}\eta^{li}  &\quad  0                                                                           \\
    0  &\quad    \frac{1}{2}\epsilon_{ab}                                                                        \\
      \end{array}
\right)\delta^{2}(u-v).
\label{inv}
\end{eqnarray}
Furthermore, the Dirac brackets among two functionals, say  $A,B$, are expressed by 
\begin{equation}
\{A(x),B(y)\}_D = \{A(x),B(y)\}_{P}\\ - \int dudv\{A(x),\chi_{\alpha}(u)\}C^{-1}_{\alpha\beta}\{\chi_{\beta}(v),B(y)\},  
\label{Dbra}
\end{equation}
where $\{A(x),B(y)\}_{P}$ is the usual Poisson bracket between the functionals $A,B$ and $\chi_{\alpha}(u),\chi_{\beta}(v)$ is the set of second class constraints. Hence, by using (\ref{inv}) and (\ref{Dbra}) we obtain the following Dirac's brackets of the theory
\begin{eqnarray}
\{e^{i}_{a}(x),e^{j}_{b} (y)\}_{D} &=& \frac{\epsilon_{ab}}{2\lambda}\eta^{ij}\delta^{2}(x-y),\nonumber  \\
\{\omega_{a} (x),\omega_{b} (y)\}_{D} &=& \frac{1}{2}\epsilon_{ab}\delta^{2}(x-y), \nonumber  \\
\{e^i_{a} (x),\Pi^{b}_{j} (y)\}_D &=& \frac{1}{2} \delta^b{_{a}} \delta^i{_{j}}\delta^{2}(x-y), \nonumber  \\
\{\Pi^a_{i} (x),\Pi^{b}_{j} (y)\}_D&=& \frac{1}{2} \epsilon^{ab} \eta_{ij}\delta^{2}(x-y), \nonumber  \\
\{P^a (x),P^{b} (y)\}_D &=& \frac{1}{2} \epsilon^{ab} \delta^{2}(x-y), \nonumber  \\
\{\omega_{a} (x),P^{b} (y)\}_{D} &=& \frac{1}{2} \delta^b{_{a}}\delta^{2}(x-y), 
\end{eqnarray}
it is important to note   that the algebra of the first class  constraints under the Dirac brackets coincide with (\ref{alg}). Moreover, we define the following gauge generator 
\begin{equation}
G=\int dx^2 \left[ \Lambda^i \gamma_{i} + \theta  \gamma \right], 
\end{equation}
thus the following gauge transformations arise 
\begin{eqnarray}
\delta e_{a}^i &=& -\frac{1}{2} \partial_a \Lambda^i - \frac{\Lambda^k}{2} \epsilon_{k}{^{i}} \omega_a+ \frac{\theta}{ 2}  \epsilon_{k}{^{i}} e^{k}_a,  \nonumber \\
\delta \omega_a &= &-\frac{1}{2} \partial_a \theta -\frac{1}{2} \Lambda^i \epsilon_{il} e^l_a,  
\end{eqnarray}
where we can observe that these gauge transformations correspond  to those  found in FJ formalism (see the section below).  In this manner, our results complete those reported in the literature \cite{13}. 
\section{ Faddeev-Jackiw  symplectic framework}
Now,  the theory  will be  analyzed by using the  FJ symplectic formalism.   For this aim, we write   the Lagrangian (\ref{lag}) in the following form 
\begin{eqnarray}
\overset{(0)}{\mathcal{L}} &=& \epsilon^{ab}\omega_{b}\dot{\omega}_{a} + \lambda\epsilon^{ab}e^{i}_{b}\dot{e}_{ai} - V^{(0)},
\label{lags}
\end{eqnarray}
where $V^{(0)}=-\omega_{0}\{2\epsilon^{ab}\partial_{a}\omega_{b} + \lambda\epsilon^{ab}\epsilon_{ij}e^{i}_{a}e^{j}_{b}\} - e^{i}_{0}\{2\lambda\epsilon^{ab}\partial_{a}e_{bi} + 2\lambda\epsilon^{ab}\epsilon_{ij}e^{j}_{a}\omega_{b}\}$ is identified as the symplectic potential. From the symplectic Lagrangian (\ref{lags}) we identify the following   symplectic variables given by  $\overset{(0)}\xi =\left(e^{i}_{a}, e^{i}_{0}, \omega_{a}, \omega_{0}\right)$ and the 1-forms $\overset{(0)}{a} = \left(\lambda\epsilon^{ab}e_{bi}, 0, \epsilon^{ab}\omega_{b}, 0 \right)$. In this manner, the symplectic matrix  given by $\overset{(0)}{f}_{ij}=\frac{\delta a_{j}}{\delta\xi^{i}} - \frac{\delta a_{i}}{\delta\xi^{j}}$ takes the form 
\begin{eqnarray}
\label{eq}
\overset{(0)}{f}_{ij}=
\left(
  \begin{array}{cccc}
   2\lambda\epsilon^{ag}\eta_{ij}  &\quad   0  &\quad\quad	  0 	&\quad\quad  0                                                                         \\
    0  &\quad    0  &\quad\quad	  0  	&\quad	\quad  0                                                                       \\
   0 &\quad  0 &\quad\quad	  2\epsilon^{ab} 	&\quad \quad	 0	\\
   0 &\quad  0 &\quad\quad	 0	&\quad\quad  0 
      \end{array}
\right) \delta^2(x-y),
\end{eqnarray}
where we can observe that $\overset{(0)}{f}_{ij}$ is singular. The null vectors of that  matrix are given by \\ $\overset{(0)}{{\mathcal{V}^i_{1}}} =\left(0, v^{e^{i}_{0}}, 0, 0\right)$ and $\overset{(0)}{\mathcal{V}^i_{2}} =\left(0, 0, 0, v^{\omega_{0}}\right)$, where $v^{e^{i}_{0}}$ and $v^{\omega_{0}}$ are arbitrary functions. Hence, from the null vectors we obtain the following FJ constraints \cite{17}
\begin{eqnarray}
\overset{(0)}{\Omega}_{i} &=& \int dx^{2}\overset{(0)}{{V^i_{1}}}\frac{\delta}{\delta\overset{(0)}{\epsilon^{i}}}\int dy^{2}\overset{(0)}{V}(\xi) = \epsilon^{ab}\partial_{a}e_{bi} + \epsilon^{ab}\epsilon_{ij}e^{j}_{a}\omega_{b} = 0, \\
\overset{(0)}{\beta} &=& \int dx^{2}\overset{(0)}{V^i_{2}}\frac{\delta}{\delta\overset{(0)}{\epsilon^{i}}}\int dy^{2}\overset{(0)}{V}(\xi) = \epsilon^{ab}\partial_{a}\omega_{b} + \frac{\lambda}{2}\epsilon^{ab}\epsilon_{ij}e^{i}_{a}e^{j}_{b} = 0,
\end{eqnarray}
we can observe that these constraints correspond to the secondary constraints obtained in Dirac's approach (see the previous section). Furthermore, we need to know  if there are more FJ constraints. Hence, we calculate the following system \cite{17, 18}
\begin{eqnarray}
\bar{f}_{kj}^{}\dot{\xi}^{(0)j}=Z_{k}^{}(\xi),
\label{p}
\end{eqnarray}
where
\begin{eqnarray}
\bar{f}_{kj}^{}=\left(
\begin{array}{cc}
f^{(0)}_{ij} \\
\frac{\delta\Omega^{(0)}_i}{\delta\xi^{(0)j}} \\
\frac{\delta\beta^{(0)}}{\delta\xi^{(0)j}}
\end{array}\right)\hphantom{111}\mathrm{and}\hphantom{111}Z_{k}^{}=
\left(
\begin{array}{cc}
\frac{\delta \mathop{\mathcal{V}}^{(0)}}{\delta\xi^{(0)j}} \\
0\\
0
\end{array}
\right), 
\end{eqnarray}
thus, we  construct the following symplectic matrix 
\begin{eqnarray}
\label{eq}
\bar{f}_{ij}=
\left(
  \begin{array}{cccc}
   2\lambda\epsilon^{ag}\eta_{il}  &\quad   0  &\quad\quad  0	 &\quad\quad 	 0                                                                         \\
    0  &\quad    0  &\quad\quad  0  &\quad\quad  0                                                                       \\
   0 &\quad  0 &\quad\quad  2\epsilon^{ag} &\quad\quad  0	\\
   0 &\quad  0 &\quad\quad  0 &\quad\quad  0 		\\
  \epsilon^{ag}\eta_{il} \partial_{a}+  \epsilon^{gb}\epsilon_{il}\omega_{b}	&\quad	0	&\quad\quad	\epsilon^{ag}\epsilon_{ij}e^{j}_{a}	&\quad\quad	0	\\
	\lambda\epsilon^{ag}\epsilon_{il}e^{i}_{a}	&\quad	0	&\quad\quad	\epsilon^{ag}\partial_{a}		&\quad\quad		0
      \end{array}
\right)\delta^{2}(x-y),
\end{eqnarray}
we observe that this matrix is not square, however has null vectors. The null vectors are  given by
$
\bar{\mathcal{V}}_{1} = \left(\delta^{l}_{k}\partial_{a}v^{k} - \epsilon^{l}_{k}\omega_{a}v^{k}, 0, \lambda\epsilon_{lj}e^{j}_{a}v^{l}, 0, -2\lambda v^{k}, 0\right) $ and $\bar{\mathcal{V}}_{2} = \left(\epsilon^{l}_{j}e^{j}_{a}v^{\lambda}, 0, -\partial_{a}v^{\lambda}, 0, 0, -2v^{\lambda}\right)$. On the other hand,  $Z_{k}$ is given by

\begin{eqnarray}
\label{eq}
\bar{Z}_{k}=
\left(
  \begin{array}{c}
   \frac{\delta\overset{(0)}{V}}{\delta\xi^{i}}	\\
0	\\
0
      \end{array}
\right)
=
 \left(
\begin{array}{c}
-2\lambda\omega_{0}\epsilon^{ab}\epsilon_{lj}e^{j}_{b} + 2\lambda\epsilon^{ba}\partial_{b}e_{0l} - 2\lambda\epsilon^{ab}\epsilon_{il}\omega_{b}e^{i}_{0} 	\\
-\overset{(0)}{\Omega_{i}}	\\
\epsilon^{ba}2\partial_{b}\omega_{0} - 2\lambda e^{i}_{0}\epsilon^{ba}\epsilon_{ij}e^{j}_{b}	\\
\overset{(0)}{\beta}	\\
0	\\	
0
\end{array}
\right).
\end{eqnarray}
The contraction of the null vectors with $\bar{Z}_{k}$ vanishes because of the constraints. For instance, from the contraction with the  first null vector we obtain 
\begin{eqnarray}
\bar{\mathcal{V}}^{\mu}_{1}\bar{Z}_{\mu} &=& 2\lambda\omega_{0}\epsilon_{lj}\{\epsilon^{ab}\partial_{a}e^{j}_{b} + \epsilon^{jk}\epsilon^{ab}e_{ka}\omega_{b}\}v^{l} + 2\lambda e^{i}_{0}\epsilon_{il}\{\epsilon^{ab}\partial_{a}\omega_{b} + \frac{\lambda}{2}\epsilon^{kj}\epsilon^{ab}e_{ak}e_{bj}\}
 \nonumber \\ &=& \omega_{0}\epsilon_{lj}\overset{(0)}{\Omega^{j}} v^l + 2\lambda e^{i}_{0}\epsilon_{il}\overset{(0)}{\beta}v^i = 0, 
\end{eqnarray}
from the contraction with the  second null vector we obtain
\begin{eqnarray}
\bar{\mathcal{V}}^{\mu}_{2}\bar{Z}_{\mu} &=& e_{0i}\epsilon^{ij}\overset{(0)}{\Omega}_{j}v^{\lambda} =0,
\end{eqnarray}
that contraction vanishes  as well. In this manner,  there are not more FJ constraints. Hence, we add the FJ constraints to the symplectic Lagrangian by using  Lagrange multipliers namely   $\alpha$,  $\zeta$,  and  the  new symplectic Lagrangian is given by 
\begin{eqnarray}
\overset{(1)}{\mathcal{L}} &=& \epsilon^{ab}\omega_{b}\dot{\omega}_{a} + \lambda\epsilon^{ab}e_{bi}\dot{e}_a^{i} - \left(\epsilon^{ab}\partial_{a}\omega_{b} + \frac{\lambda}{2}\epsilon^{ab}\epsilon_{ij}e^{i}_{a}e^{j}_{b}\right)\dot{\alpha} - \left(\epsilon^{ab}\partial_{a}e_{bi} + \epsilon^{ab}\epsilon_{ij}e^{j}_{a}\omega_{b}\right)\dot{\zeta}^{i} - \overset{(1)}{V},
\label{eq28}
\end{eqnarray}
where $\overset{(1)}{V} = \overset{(0)}{V}|_{{\overset{(0)}{\Omega}_{i}}, \overset{(0)}{\beta}} = 0$ vanishes  because of the general covariance of the theory. We can observe that $\dot{\alpha} = \omega_{0}$ and $\dot{\zeta} = e^{i}_{0}$ has been taken into the account.
From the symplectic Lagrangian (\ref{eq28}) we identify the following symplectic variables $\overset{(1)}{\xi} = \left(e^{i}_{a}, \zeta^{i}, \omega_{a}, \alpha \right)$  and the 1-forms $\overset{(1)}{a} =\left(\lambda\epsilon^{ab}e_{bi}, - (\epsilon^{ab}\partial_{a}e_{bi} + \epsilon^{ab}\epsilon_{ij}e^{j}_{a}\omega_{b}), \epsilon^{ab}\omega_{b}, -(\epsilon^{ab}\partial_{a}\omega_{b} + \frac{\lambda}{2}\epsilon^{ab}\epsilon_{ij}e^{i}_{a}e^{j}_{b})\right)$, where the new symplectic matrix has the following form
\begin{eqnarray}
\label{eq}
\overset{(1)}{f}_{ij}=
\left(
  \begin{array}{cccc}
   2\lambda\epsilon^{ag}\eta_{ij}  	&\quad   	-\epsilon^{ag}\eta_{il}\partial_{a} - \epsilon^{gb}\epsilon_{il}\omega_{b}  &\quad\quad		 0	 &\quad\quad	-\lambda\epsilon^{ag}\epsilon_{il}e^{i}_{a}                                                                 \\
    \epsilon^{ag}\eta_{il}\partial_{a} + \epsilon^{gb}\epsilon_{il}\omega_{b}	 &\quad    0  &\quad\quad  \epsilon^{ag}\epsilon_{ij}e^{j}_{a}	  &\quad\quad	  0                                                                       \\
   0 &\quad  	-\epsilon^{ag}\epsilon_{ij}e^{j}_{a}	&\quad\quad		  2\epsilon^{ag} 	&\quad\quad	-\epsilon^{ag}\partial_{a}		\\
   \lambda\epsilon^{ag}\epsilon_{il}e^{i}_{a}		&\quad  	0	 &\quad\quad		  \epsilon^{ag}\partial_{a}	 &\quad\quad	  0 	
      \end{array}
\right)\delta^{2}(x-y),
\label{eq30}
\end{eqnarray}
we can observe that the matrix is singular. In fact, this means that the system has a gauge symmetry and   it is well-known that the null vectors of the matrix (\ref{eq30}) are  generators of that symmetry  \cite{16}. In fact, the null vectors of the matrix (\ref{eq30}) are given by 
\begin{eqnarray}
\Gamma_1 &=& \left(\frac{\theta}{ 2}  \epsilon_{k}{^{i}} e^{k}_a, 0, -\frac{1}{2} \partial_a \theta, - \theta \right), \nonumber\\
\Gamma_2 &=& \left( -\frac{1}{2} \partial_a \Lambda^i - \frac{\Lambda^k}{2} \epsilon_{k}{^{i}} \omega_a,  - \lambda \Lambda^i, -\frac{1}{2} \Lambda^i \epsilon_{il} e^l_a, 0 \right), 
\end{eqnarray}
where $\Lambda^i$ and $\theta$ are gauge parameters. By using these null vectors, we find the following gauge transformations of the theory
\begin{eqnarray}
\delta e_{a}^i &=& -\frac{1}{2} \partial_a \Lambda^i - \frac{\Lambda^k}{2} \epsilon_{k}{^{i}} \omega_a+ \frac{\theta}{ 2}  \epsilon_{k}{^{i}} e^{k}_a,  \nonumber \\
\delta \omega_a &= &-\frac{1}{2} \partial_a \theta -\frac{1}{2} \Lambda^i \epsilon_{il} e^l_a, 
\end{eqnarray}
where we can observe that these transformations coincide with those found in the Dirac scheme. Furthermore,we have commented above that  Hussain's theory is diffeomorphism covariant and a (analog) Hamiltonian constraint is not present in the theory \cite{13}. In fact, we can attend   those points by redefining the gauge parameters as $\Lambda^i= 2e^i_a \tau^a $ and $\theta=- 2\omega_a \tau^a $, hence the gauge transformations take the form 
\begin{eqnarray}
\delta e^i_a&=& \mathcal{L}_\tau e^i_a + \tau^b \left[ \partial_a e_b ^i- \partial_b e_a^i  \right] + \tau^b \epsilon{^{i}}_ k \left[  e_a^k \omega_b - e_b^k \omega_a\right], \nonumber \\
\delta \omega_a &=& \mathcal{L}_\tau \omega_a + \tau^b \left[ \partial_a \omega_b - \partial_b \omega_a+ \tau^b \epsilon _{il} e^i_b e^l _a\right]  , 
\end{eqnarray}
which  correspond (on shell) to diffeomorphisms and it is  an internal symmetry of the theory. In this manner, we have reproduced by other way the results reported in \cite{13}.     \\
On the other hand,  we have showed that there are not more FJ constraints and the theory has a gauge symmetry, therefore, in order to obtain a symplectic tensor  we fixing  the temporal gauge
\begin{eqnarray}
e^{i}_{0} &=& 0,	 \nonumber  \\
\omega_{0} &=& 0, 
\end{eqnarray}
this implies  that ${\alpha}=cte , {\zeta}^{i}=cte$. In this manner, by adding the temporal gauge as constraints,  the new symplectic Lagrangian is given by
\begin{eqnarray}
\overset{(2)}{\mathcal{L}} &=& \epsilon^{ab}\omega_{b}\dot{\omega}_{a} + \lambda\epsilon^{ab}e_{bj}\dot{e}^{i}_{a} - \left(\epsilon^{ab}\partial_{a}\omega_{b} + \frac{\lambda}{2}\epsilon^{ab}\epsilon_{ij}e^{i}_{a}e^{j}_{b} - \rho\right)\dot{\alpha} - \left(\epsilon^{ab}\partial_{a}e_{bi} + \epsilon^{ab}\epsilon_{ij}e^{i}_{a}\omega_{b} - \sigma_{i}\right)\dot{\zeta}^{i}, \nonumber \\
\end{eqnarray}
where we  choose  the following symplectic variables $\overset{(2)}{\xi}=\left(e^{i}_{a}, \zeta^{i}, \omega_{a}, \alpha, \rho, \sigma_{i}\right)$ and the 1-forms $\overset{(2)}{a}=\left(\lambda\epsilon^{ab}e_{bi}, -(\epsilon^{ab}\partial_{a}e_{bi} + \epsilon^{ab}\epsilon_{ij}e^{i}_{a}\omega_{b} - \sigma_{i}), \epsilon^{ab}\omega_{b}, -(\epsilon^{ab}\partial_{a}\omega_{b} + \frac{\lambda}{2}\epsilon^{ab}\epsilon_{ij}e^{i}_{a}e^{j}_{b} - \rho), 0, 0\right)$. Now, the symplectic matrix takes the following form

\begin{eqnarray}
\label{eq}
\overset{(2)}{f}_{ij}=
\left(
  \begin{array}{cccccc}
   2\lambda\epsilon^{ag}\eta_{il} 	 &\quad   -\epsilon^{ag}\eta_{il}\partial_{a} - \epsilon^{gb}\epsilon_{il}\omega_{b}  &\quad 	0	 &\quad	-\lambda\epsilon^{ag}\epsilon_{il}e^{i}_{a}		&\quad\quad		0	&\quad\quad		0                                                                 \\
    \epsilon^{ag}\eta_{il}\partial_{a} + \epsilon^{gb}\epsilon_{il}\omega_{b}	 &\quad    0	  &\quad   \epsilon^{ag}\epsilon_{ij}e^{j}_{a}	  &\quad  0		&\quad\quad		0	&\quad\quad		-\delta^{i}_{j}	                                                                       \\
   0 	&\quad  	-\epsilon^{ag}\epsilon_{ij}e^{j}_{a}	&\quad 	 2\epsilon^{ag} 	&\quad	-\epsilon^{ag}\partial_{a}		&\quad\quad		0	&\quad\quad		0
\\
   \lambda\epsilon^{ag}\epsilon_{il}e^{i}_{a}		&\quad  	0	 &\quad	  \epsilon^{ag}\partial_{a}	 &\quad  0 	&\quad\quad		-1	&\quad\quad		0
\\
0	&\quad	0	&\quad	0	&\quad	1	&\quad\quad		0	&\quad\quad		0
\\
0	&\quad	\delta^{i}_{j}	&\quad	0	&\quad	0	&\quad\quad		0	&\quad\quad		0	
      \end{array}
\right) \nonumber  \\  \times \delta^2(x-y), \nonumber  \\
\end{eqnarray}
we observe that this matrix is a symplectic tensor, and its inverse is given by
\begin{eqnarray}
\label{eq}
\overset{(2)}{f_{ij}}^{-1}=
\left(
  \begin{array}{cccccc}
   
\frac{1}{2\lambda}\epsilon^{ag}\eta^{ij} 	 &\quad   0		&\quad\quad		0	 &\quad\quad	0		&\quad\quad		-\frac{\epsilon^{i}{_{j}}}{2}e^{j}_{a}		&\quad\quad		\frac{1}{2\lambda}\left(-\delta^{i}_{j}\partial_{a} + \epsilon^{i}_{j}\omega_{a}\right)
\\
   0	 &\quad    0	 	&\quad\quad		 0	  &\quad\quad	  0		&\quad\quad		0	&\quad\quad		\delta^{i}_{j}	 \\
   0 	&\quad  	0	&\quad\quad		 \frac{1}{2}\epsilon_{ab}	 	&\quad\quad		0	&\quad\quad		-\frac{1}{2}\partial_{a}	&\quad\quad		\frac{1}{2}\epsilon_{ij}e^{j}_{a}
\\
   0	&\quad  	0	 &\quad\quad		0	 &\quad\quad		 0 	&\quad\quad		1	&\quad\quad		0
\\
\frac{\epsilon^{i}{_{j}}}{2}e^{j}_{a}		&\quad	0	&\quad\quad		\frac{1}{2}\partial_{a}	&\quad\quad		-1	&\quad\quad		0	&\quad\quad		\epsilon_{ij}\frac{\epsilon^{ab}}{2}\partial_{a}e^{j}_{b}
\\
\frac{1}{2\lambda}\left(\delta^{i}_{j}\partial_{a} - \epsilon^{i}{_{j}}\omega_{a}\right)		&\quad	-\delta^{i}_{j}	&\quad\quad		-\frac{1}{2}\epsilon_{ij}e^{j}_{a}	&\quad\quad		0	&\quad\quad		-\frac{\epsilon_{ij}\epsilon^{ab}}{2}\partial_{a}e^{j}_{b}		&\quad\quad		0	
      \end{array}
\right)  \delta^2(x-y).  \nonumber \\
\label{symf}
\end{eqnarray}
Therefore, from the symplectic tensor (\ref{symf}) we can identify the generalized FJ brackets by means of 
\begin{eqnarray}
\{\xi_{i}^{(2)}(x),\xi_{j}^{(2)}(y)\}_{FD}=[f^{(2)}_{ij}(x,y)]^{-1},
\end{eqnarray}
thus, the following generalized brackets arise 
\begin{eqnarray}
\{e^{i}_{a},e^{j}_{b}\}_{FJ} &=& \frac{1}{2\lambda}\epsilon_{ab}\eta^{ij}\delta^{2}(x-y), \nonumber  \\
\{\omega_{a},\omega_{b}\}_{FJ} &=& \frac{1}{2}\epsilon_{ab}\delta^{2}(x-y), 
\end{eqnarray}
we can observe that the Dirac brackets and the FJ ones coincide to each other. 
\section{ Summary and conclusions}
In this paper a detailed   canonical and symplectic analysis for Husain's gravity has been performed. The complete structure of the Dirac  constraints and the  algebra between them  has been reported,  we eliminated the second class constraints by introducing the Dirac brackets, then  have used a temporal gauge in order to construct the new Dirac's brackets. Furthermore, with respect to the symplectic method, we obtained the complete set of FJ constraints, the gauge transformations were found and the diffeomorphisms were reported as a internal symmetry of the theory, then   by fixing the temporal gauge  a symplectic tensor has been constructed. From the symplectic tensor  the generalized FJ brackets were identified and we showed that Dirac's and FJ brackets coincide to each other. It is important to comment that in Dirac's formulation the classification of  the constraints in first  class and second class is a difficult task, in FJ approach, however, the identification of the constraints is less complicated  and there are present less constraints  than Dirac's method. In this sense, the FJ formulation is more elegant and  economical.   
\newline
\newline
\newline
\noindent \textbf{Acknowledgements}\\[1ex]
This work was supported by CONACyT under Grant No.CB-$2014$-$01/ 240781$. We would like to  R. Cartas-Fuentevilla for discussion on the subject and reading of the manuscript.

\end{document}